\documentclass{article}
\usepackage[a4paper]{geometry}
\usepackage{comment}
\usepackage{epsfig}
\usepackage{multicol}
\usepackage{amsmath}
\usepackage{floatrow}
\usepackage[label font=bf,labelformat=simple]{subfig}% <-- changed
\usepackage{caption}
\usepackage{natbib}
\usepackage{comment}

\title{GLOBAL LINEAR STABILITY ANALYSIS OF A SLIT FLAME SUBJECT TO INTRINSIC THERMOACOUSTIC INSTABILITY}

\usepackage{array,makecell}
\usepackage{caption}
\author{Grégoire Varillon, Philipp Brokof and Wolfgang Polifke\footnote{{\small \textit{Technical University of Munich, Germany; School of Engineering and Design, Department of Engineering Physics and Computation, e-mail: gregoire.varillon@tum.de}}}}
\newcommand{\rout}{\mathcal{R}_\text{out}}
\newcommand{\rin}{\mathcal{R}_\text{in}}
\date{\today}
\begin{document}

\maketitle
\begin{abstract}
The present study makes use of the adjoint modes of the Linearized Reactive Flow (LRF) equations to investigate the Intrinsic Thermoacoustic (ITA) feedback loop of a laminar premixed slit flame.
The analysis shows that the ITA feedback loop is closed by vorticity generated in the boundary layer of the slit by impinging acoustic waves penetrating the slit.
In this region, adjoint vorticity shows a high sensitivity of the flow.
It is also hypothesised that the ITA eigenmode smoothly transitions to a purely hydrodynamic mode -- vortex shedding -- for a passive flame.
The computational domain is chosen sufficiently short so as to isolate the ITA feedback loop from cavity modes.
This analysis is made possible by the holistic character of the LRF model, i.e. a direct linearization of the non-linear reactive flow equations, including explicit finite rate chemistry and avoiding idealization of the flow.
\end{abstract}

\section{Introduction}\label{sec:intro}
Combustion devices, such as gas turbines, are prone to thermoacoustic instabilities, especially when these power generation units are fired in the lean premixed regime to limit NOx emissions \citep{LieuwYang05}.
Thermoacoustic instabilities result from the interplay of the fluctuating heat release rate with the acoustic waves developing in the compressible medium enclosed by the combustion device.
Intrinsic thermoacoustic (ITA) modes are thermoacoustic fluctuations that are not coupled with the natural acoustic frequencies of the combustion chamber or any other part of the combustion system.
In other words, they may occur even in a purely anechoic environment.
In the case of a not-fully anechoic configuration, the resulting thermoacoustic fluctuations will follow from the interaction of ITA and cavity modes (natural frequencies) of the combustion device.
ITA modes have been conceptualized from experimental results in \citep{HoeijKorni14,EmmerBombe15,BombeEmmer15} and investigated with CFD in \citep{SilvaEmmer15,CourtSelle15}.
The ITA feedback loop was then successfully formalised in \citep{BombeEmmer15} with the help of low-order models: a deterministic operator-based approach build on networks of 1D linear acoustic elements at zero Mach, connected to an acoustically compact flame described by an \textit{ad-hoc} flame transfer function (FTF).
By allowing parametric studies at low numerical costs on laminar flames, but also more complex systems such as can-annular combustor \citep{FournHaeri20a,BuschMensa20}, network models led to the categorization of thermoacoustic fluctuations into modes of acoustic and ITA origin \citep{MensaMagri18b,SilvaYong18a}.
Recently, Yong et al. \citep{YongSilva22} proposed a new paradigm for classification of thermoacoustic modes, showing that acoustic modes can transition continuously to ITA modes, and vice versa.
Network models are also used in the perspective of controlling thermoacoustic instabilities \citep{ghani_control_2021,SchaeMagri21a}
As recently reviewed in \citep{Silva23}, the understanding of the ITA feedback loop and its interaction with the cavity modes of the combuster remains to some extent an open question, in particular with regard to the stabilization of thermoacoustic modes of ITA origin.

Although they allow for a broad characterisation of thermoacoustic modes, network models brought little physical insight regarding the flow-flame interactions, which may be an essential part of the ITA feedback loop. Hence, remaining open questions are: How do the acoustics interact with the flow at the base of the flame and feed back to the flame motion?
Why do ITA modes turn unstable in some environments?
This is partially due to the fact that network models describe only 1D acoustics and no peculiarity of the flow, e.g. shear layers, stratification, recirculation zones, etc.
All these elements are included in the \textit{ad-hoc} FTF as a black-box. 
On the other hand, the data-driven approach relies on CFD simulation or experiments, where such peculiarities are described.
However, CFD simulations do not allow stability analysis as does the operator-based approach.

The Linearized Reactive Flow model provides a holistic operator-based approach to carry out a stability analysis on a realistic reactive flow, without the need for any external \textit{ad-hoc} FTF \citep{AvdonMeind19}.
LRF has already been demonstrated successful regarding thermoacoustic systems to accurately identify least stable modes \citep{AvdonMeind19,wang_global_2022,wang_linear_2022}, and proved to be superior to models using an \textit{ad-hoc} flame model \citep{MeindSilva20}.
Throughout this article, we refer to the \textit{ITA mode} as the unstable oscillations of a flame in an anechoic environment, and to the \textit{ITA loop} or \textit{ITA feedback loop} as the feedback loop originally sketched in \citep{BombeEmmer15}.
The ITA feedback loop exists in both echoic and anechoic configuration.
This work intends to bring a deeper understanding of the flow-acoustic-flame interactions of a laminar slit flame, and more specifically, of why the ITA mode may turn unstable in anechoic environments. Here, we study the slit flame configuration of \citep{SilvaEmmer15}. The computational domain of the combustion chamber is chosen sufficiently short to not sustain any cavity mode in the frequency range of interest of the present study, so as to isolate the ITA loop.
% ITA feedback loop and its connection to the flow characteristics of the slit.
The rest of the article is organized as follow: the configuration and modelling choices are presented in Sec.~\ref{sec:model}, the LRF equations are recalled in Sec.~\ref{sec:linearized_equations} along with validation against FTF from transient CFD simulations. Stability results are presented and discussed in Sec.~\ref{sec:results}, where the physics of the ITA loop is investigated through direct and adjoint eigenmodes.
The results are summarised in Sec.~\ref{sec:conclusion},

\section{Modelling}\label{sec:model}
\begin{figure}[ht]
        \centering
        \sidesubfloat[]{\includegraphics[width=0.18\textwidth]{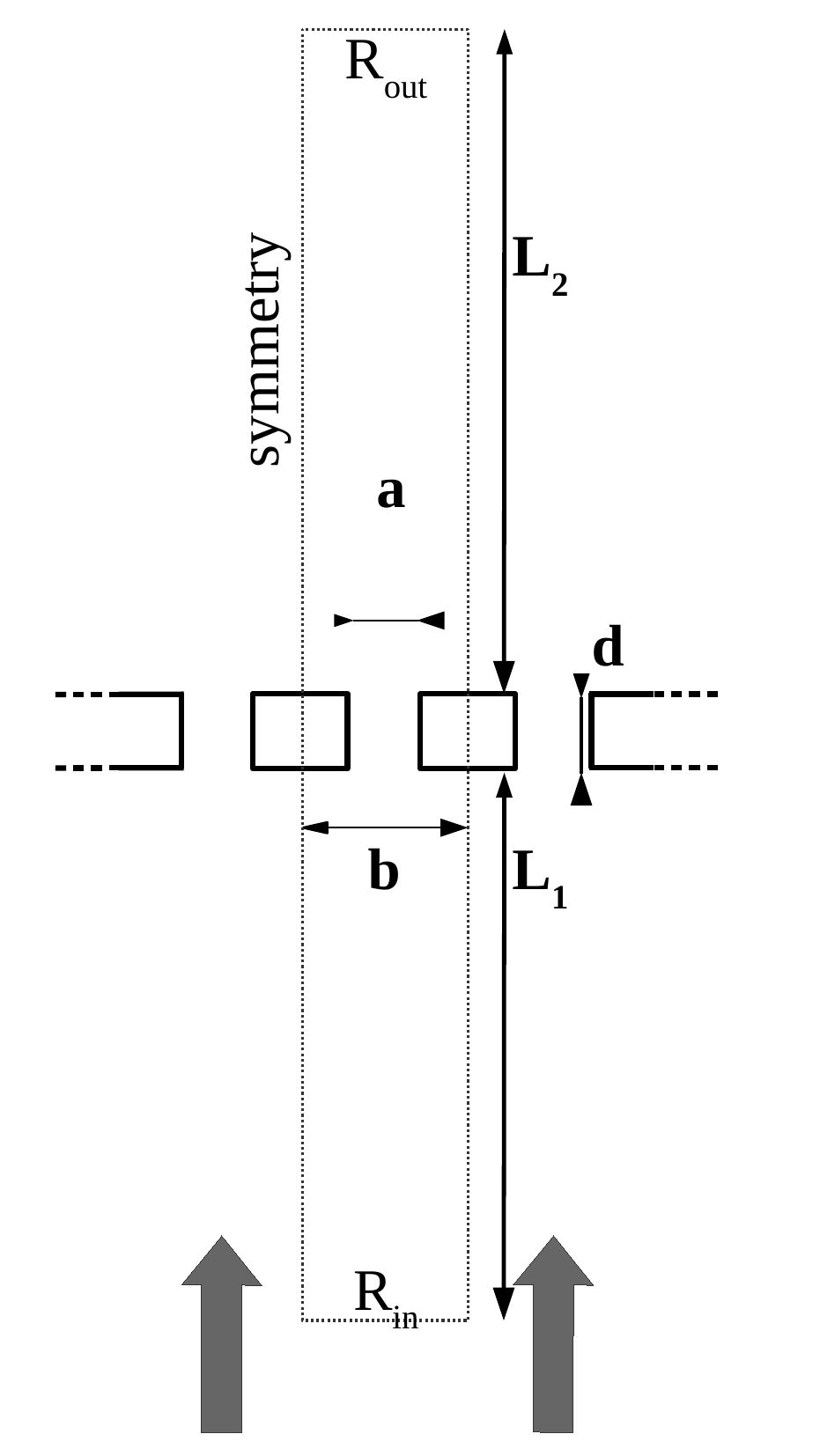}}
        \sidesubfloat[]{\includegraphics[width=0.3\textwidth]{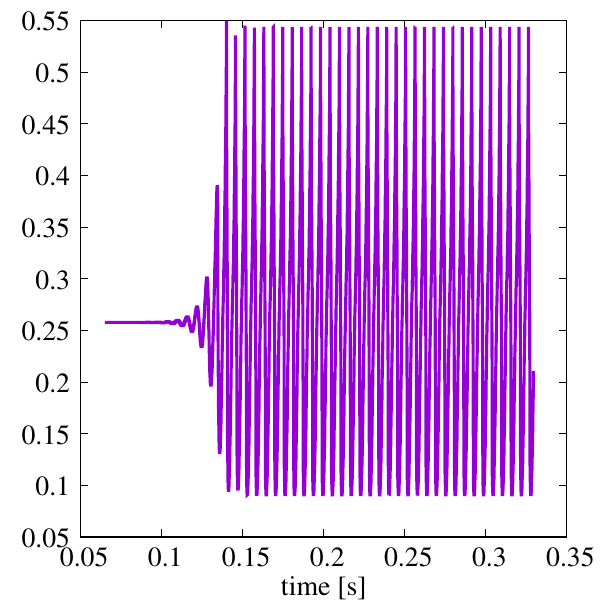}}
        \sidesubfloat[]{\includegraphics[width=0.3\textwidth]{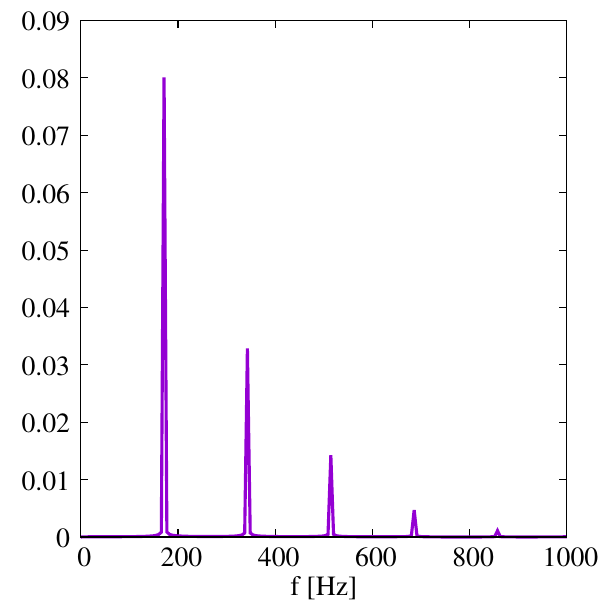}}
        \caption{Configuration of the slit flame (a). Global heat release rate vs. time (b) and amplitude of Fourier coefficients of the global heat release rate fluctuation (c), both obtained from CFD.}
        \label{fig:limit_cycle}
\end{figure}

The configuration of interest stems from the Kornilov test-rig \citep{KorniRook09} considered in 2D (Fig.~\ref{fig:limit_cycle}a).
A fresh premixture of air and methane is injected in an array of slits.
The system is assumed to be invariant in the direction of the slit length and periodic in the direction of the slit width.
Such a system is modelled as a reacting perfect gas of density $\rho$, sensible enthalpy $h$, and pressure $p$.
Velocity components are denoted $u_i$.
The evolution equation writes \citep{PoinsVeyna12a}:
\begin{subequations}
    \begin{align}
           & \partial_t \rho + \partial_j \rho u_j = 0, \\
           & \partial_t \rho u_i + \partial_j \rho u_i u_j + \partial_i p = \partial_j \tau_{ij}, \\
           & \partial_t (\rho h - p) + \partial_j \rho u_j h = \partial_j (\alpha \partial_j h) + \dot{\omega}_T, \\
           & \partial_\rho Y_k + \partial_j \rho u_j Y_k = \partial_j (D_k\partial_j Y_k) + \dot{\omega}_k.
    \end{align}
    \label{eq:mean_flow}
    \end{subequations}
The viscous stress $\tau_{ij}$ is linked to strain \textit{via} Stokes' hypothesis 
\begin{equation}
    \tau_{ij}=\mu (\partial_j u_i + \partial_i u_j - \dfrac{2}{3}\partial_j u_k \delta_{ij}),
\end{equation}
where the dynamic viscosity $\mu$ is obtained from Sutherland's law.
Heat diffusion is modelled with Fourier's law, with a heat diffusivity $\alpha$.
The combustion of methane is modelled by the two-steps BFER mechanism \citep{BibrzPoins10}, with species reaction rate $\dot{\omega}_k$ and heat release rate $\dot{\omega}_T$.
Species are described by their respective masse fraction $Y_k$, molecular diffusion is described with Fick's law, with diffusivity $D_k$.

The base-flow is characterized by an inlet velocity $u_\text{in}=0.4m/s$ at temperature $T_\text{w}=373K$ and with an equivalence ratio $\Phi=0.8$.
The perforated plate is considered isothermal at $373K$ from experiments \citep{KorniRook09}.
Such a configuration with anechoic inlet and outlet is ITA unstable, with a limit cycle at 173.5 Hz (Fig.~\ref{fig:limit_cycle}), as previously shown in \citep{SilvaEmmer15}.

\section{Linearized equations}
\label{sec:linearized_equations}
\begin{figure}[ht]
        \centering
        \includegraphics[width=.65\textwidth]{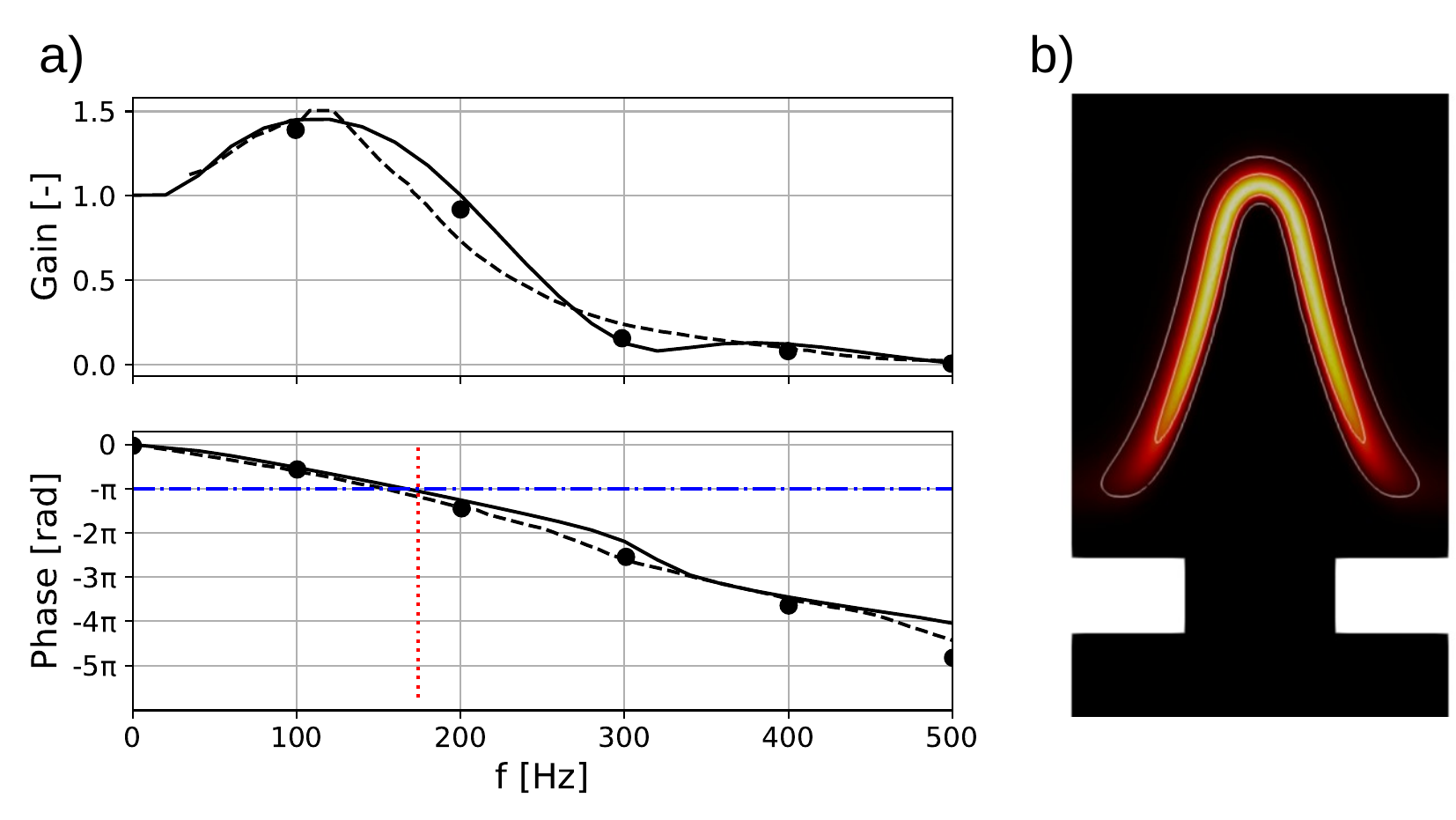}
        \caption{(a) Flame transfer function obtained from LRF (solid), from the experiments of Kornilov \citep{KorniRook09} (dashed) and the DNS of Duchaine \citep{DuchaBoudy11} (dots). Blue dash-dotted horizontal curve at $-\pi$ and vertical red dotted curve at the ITA limit cycle frequency $173.5Hz$.(b) Heat release rate $\bar{\dot{\omega}}_T$ of the base-flow from CFD.}
        \label{fig:dQMean}
\end{figure}

Each flow variable is decomposed as $q=\bar{q} + \epsilon q'$, $\epsilon\ll 1$, with $\bar{q}$ the steady base-flow and $q'$ the linear perturbations.
The baseflow $\bar{q}$ is a steady solution of \eqref{eq:mean_flow} (Fig.~\ref{fig:dQMean}b).
Such a solution is obtained by taking the low-Mach limit of \eqref{eq:mean_flow}, therefore suppressing the acoustic feedback, which is known to be stable \citep{Silva23}.
The evolution equations for $q'$ are obtained by linearizing \eqref{eq:mean_flow} around $\bar{q}$, here reproduced from \citep{AvdonMeind19},
\begin{subequations}
    \begin{align}
       & \partial_t \rho' + \partial_j (\Bar{\rho}u_j'+\rho'\Bar{u}_j)=0, \quad \partial_t(\Bar{\rho} u_i' + \rho'\Bar{u}_i)+\partial_j(\Bar{\rho}\Bar{u}_iu_j' + \Bar{\rho}\Bar{u}_j u_i'+\Bar{u}_j\Bar{u}_i\rho') = -\partial_ip' + \partial_j \tau_{ij}', \\
        & \partial_t(\Bar{\rho}h' + \rho'\Bar{h}-p') + \partial_j(\Bar{\rho}\Bar{u}_jh' + \Bar{\rho}\Bar{h}u_j' + \Bar{h}\Bar{u}_j \rho')=\partial_j(\Bar{\alpha}\partial_jh') + \dot{\omega}_T', \\
        & \partial_t(\Bar{\rho}Y_k'+\rho'\Bar{Y}_k) + \partial_j(\Bar{\rho}\Bar{u}_jY_k'+\Bar{\rho}\Bar{Y}_k u_j' + \Bar{Y}_k\Bar{u}_j \rho') = \partial_j(\Bar{D}\partial_j Y_k) + \dot{\omega}_k'.
        \end{align}
    \label{eq:linearized_eq}
\end{subequations}
The linearization of the reaction rates is given in App.~A of \citep{MeindSilva20}.
The heat release rate perturbation $\dot{\omega}'_T$ is reformulated to integrate an adjustable \textit{ad-hoc} prefactor $\Gamma$ that will be used later for parametric studies:
\begin{equation}
    \dot{\omega}'_T=\Gamma\sum_{n=1}^2 \Delta h^0_n \mathcal{Q}'_n,
    \label{eq:prefactor_hrr}
\end{equation}
with $\mathcal{Q}'_n$ the perturbation of the reaction progress rate.
The evolution equations for linear perturbations \eqref{eq:linearized_eq} are discretized in space with the DG-FEM \citep{MeindAlbay20}.
Boundary conditions are obtained by linearizing boundary conditions of the base-flow and imposed weakly through fluxes in the DG-FEM formulation.
At the inlet and outlet, boundary conditions are imposed such as to control the reflection or transmission of acoustic waves normal to these boundaries through the acoustic characteristic waves $\hat{f}$ and $\hat{g}$ defined as:
\begin{equation}
    \hat{f}=\dfrac{\hat{p}}{\Bar{\rho}\Bar{c}} + \hat{u}_x \quad ; \quad \hat{g}=\dfrac{\hat{p}}{\Bar{\rho}\Bar{c}} - \hat{u}_x, \quad \rin = \dfrac{\hat{f}_\text{in}}{\hat{g}_\text{in}} \quad ; \quad \rout = \dfrac{\hat{g}_\text{out}}{\hat{f}_\text{out}}.
    \label{eq:definition_rout}
\end{equation}
where $\hat{q}$ is the complex Fourier amplitude to $q'$ and $\rin$ and $\rout$ are the reflection coefficients.
%The reflection coefficients  are defined as
%\begin{equation}
%    \rin = \dfrac{\hat{f}_\text{in}}{\hat{g}_\text{in}} \quad ; \quad \rout = \dfrac{\hat{g}_\text{out}}{\hat{f}_\text{out}}.
%    \label{eq:definition_rout}
%\end{equation}
Setting $\rout = -1$ or $\rout = 1$ corresponds to an open-end or a hard wall outlet, respectively.
Equation \eqref{eq:linearized_eq} together with boundary conditions is discretized on a computational domain of dimensions $L_1=14$mm, $L_2=5$mm, $d=1$mm, $b=5$mm and $a=2$mm, composed of $24\,10^3$ discontinuous finite elements of order 1, leading to $6.3 \times 10^5$ degrees of freedom.
The space-discrete and time-continuous problem is matrix shaped as
    \begin{equation}
        \mathcal{M}\partial_t\mathbf{q}=\mathcal{K}\mathbf{q} + \mathbf{b},
        \label{eq:perturbations}
    \end{equation}
with $\mathbf{q}$ the vector of discrete state variables, $\mathbf{b}$ an external forcing and $\mathcal{M}$ and $\mathcal{K}$ the mass and stiffness matrices, respectively.

The flame transfer function (FTF) relates fluctuations of the heat release rate to fluctuations of the inlet velocity.
The FTF is expressed from the linearized equations by considering the temporal Fourier transform of \eqref{eq:linearized_eq},
\begin{equation}
    \text{FTF}=\dfrac{\hat{\dot{\omega}}_T/\bar{\dot{\omega}}_T}{\hat{u}_\text{in}/\Bar{u}_\text{in}}.
\end{equation}
The FTF is computed with LRF from \eqref{eq:perturbations} with $\mathbf{f}$ as an axial velocity forcing at the inlet (Fig.~\ref{fig:dQMean}a).
The FTF from LRF agrees with the experiments of \citep{KorniRook09} and the DNS computations \citep{DuchaBoudy11}, which validates the LRF approach to capture the flame dynamics.
In particular, the low-frequency limit, the low-pass filter behaviour and the excess gain are recovered with reasonable accuracy.
The FTF also shows that the $\pi$ criterion is matched at the oscillation frequency of the limit cycle (Fig.~\ref{fig:limit_cycle}), i.e. at that frequency the heat release rate and pressure fluctuations are in phase, yielding a positive Rayleigh index \citep{Cande02}.

\section{Global stability analysis}\label{sec:results}
\begin{figure}[ht]
        \centering
        \sidesubfloat[]{\includegraphics[width=0.33\textwidth]{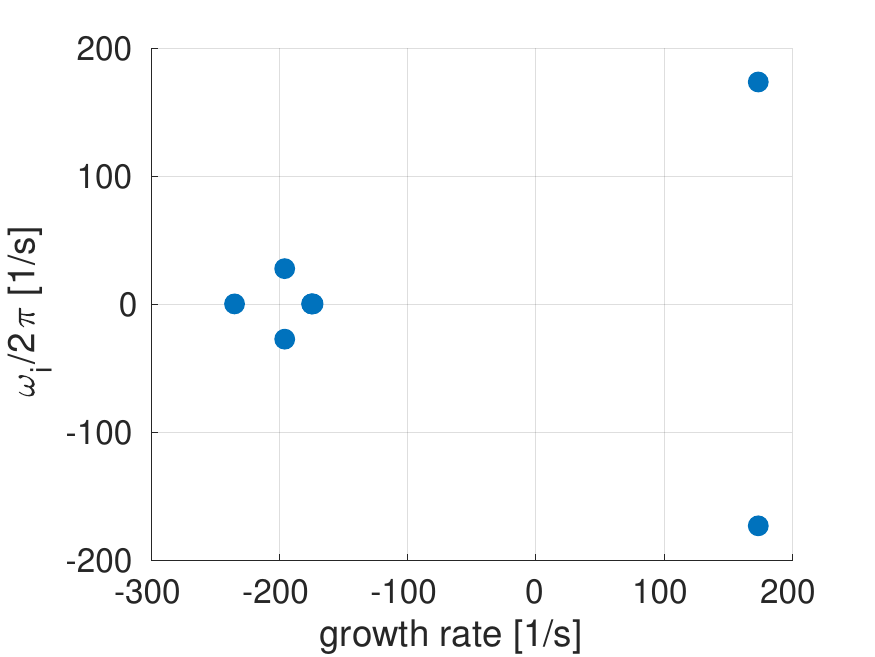}}
        \sidesubfloat[]{\includegraphics[width=0.33\textwidth]{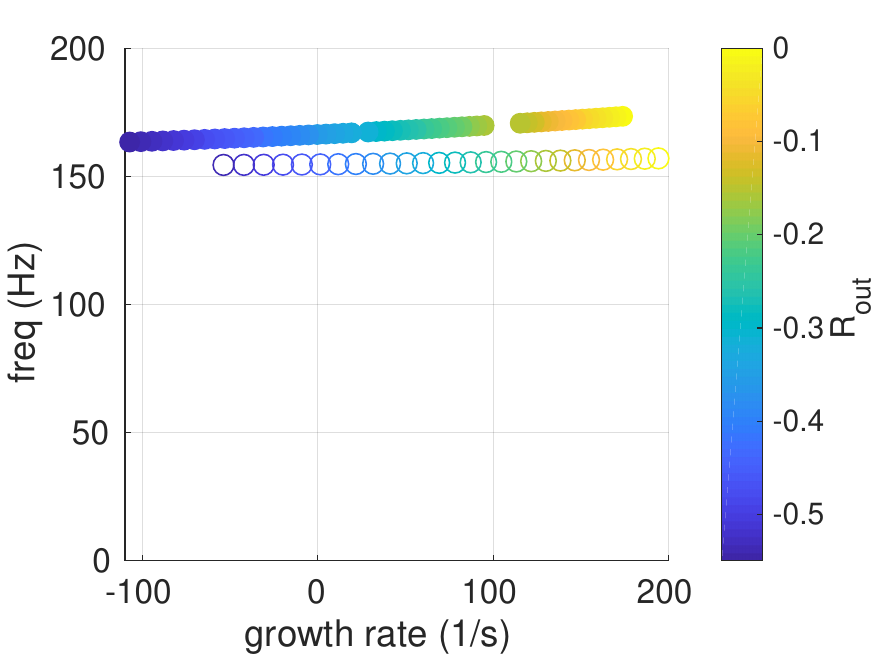}}
        \sidesubfloat[]{\includegraphics[width=0.33\textwidth]{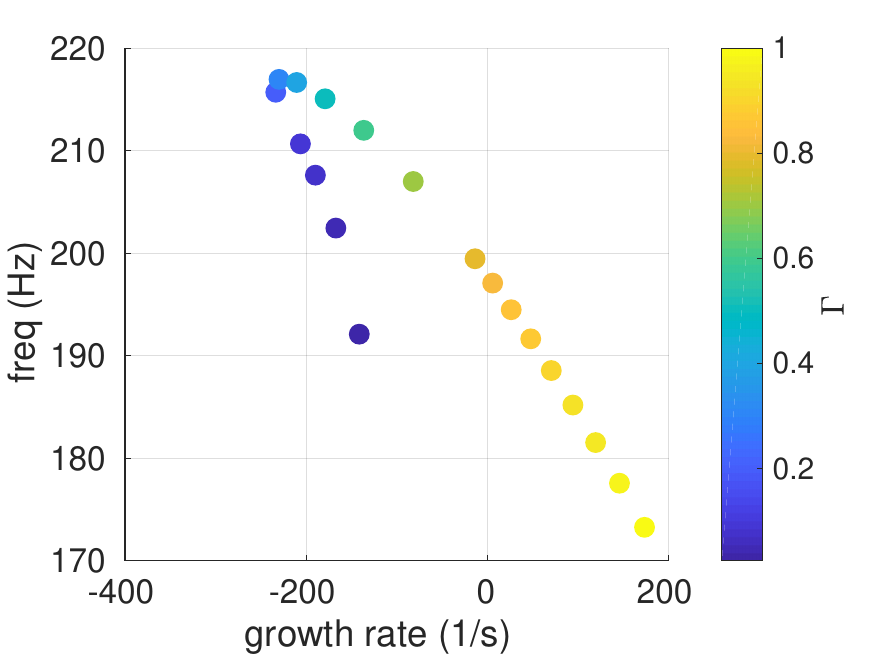}}
        \caption{(a) Stability map for $\rout=0$, $\Gamma=1$ from LRF, and least stable eigenvalue (b) for varying $\rout$, $\Gamma=1$, from LRF (filled) and Low-order model (circle) and (d) for varying $\Gamma$, $\rout=0$, from LRF.
        Parameters $\rout$ and $\Gamma$ are defined in Eqs.~\eqref{eq:definition_rout} and \eqref{eq:prefactor_hrr}, respectively.}
        \label{fig:sweep_r}
\end{figure}

\begin{figure}[ht]
    \centering
    \includegraphics[width=0.7\linewidth]{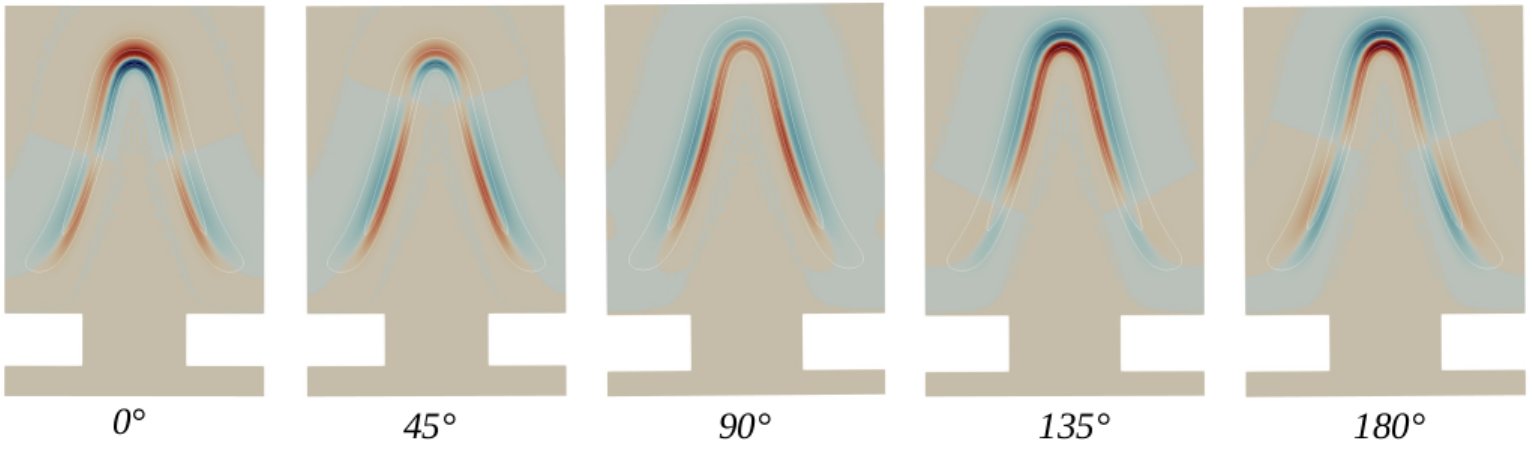}\par
\caption{Linear response of the heat-release rate $\dot{\omega}'_T$ from LRF of the ITA eigenmode, at various phase angle.}
\label{fig:dQ_LRF}
\end{figure}

Equation \eqref{eq:perturbations} for $b=0$ is Laplace-transformed, leading to the eigenvalue problem
\begin{equation}
        \lambda \mathcal{M}\widetilde{\mathbf{q}}=\mathcal{K}\widetilde{\mathbf{q}}.
        \label{eq:evp}
\end{equation}
The real and imaginary part of the eigenvalue $\lambda$ represent the growth rate and pseudo-frequency of the associated eigenmode $\widetilde{\mathbf{q}}$, respectively.
This generalized eigenvalue problem is solved with the SLEPc library, using the shift-and-invert transform \citep{HernaRoman05}.
%\begin{figure}[ht]
%        \centering
%        \includegraphics[width=0.3\textwidth]{Figs/stability_map_r0.pdf}
%        \caption{Stability map from LRF with non-reflecting outlet.}
%        \label{fig:stability_map_r0}
%\end{figure}
The stability map for $\rout=0$ (Fig.~\ref{fig:sweep_r}a) shows a least stable eigenpair with positive growth rate at 174 Hz.
This denotes that the baseflow is unstable to compressible fluctuations, and that only one feedback loop drives the instability.
The eigenfrequency of this unstable mode closely matches the frequency of the limit cycle identified from CFD (174Hz, Fig.~\ref{fig:limit_cycle}), also matching the $\pi$ criterion.
Qualitatively, the heat release rate fluctuation corresponding to this unstable eigenmode is a wrinkle of the flame sheet propagating along the length of the flame (Fig.~\ref{fig:dQ_LRF}), similar to the nonlinear limit cycle.
This least stable eigenvalue is the \textit{ITA eigenmode} associated with the ITA loop.
Similar to the ITA loop, the ITA eigenmode exists in both, echoic and anechoic, environment.

\subsection{From non-reflective to open-end combustion chamber}
As observed in experiments and predicted by CFD simulations \citep{Silva23}, the ITA eigenmode becomes stable when the outlet is changed from non-reflecting to open-end, i.e. $\rout$ is sweeped from 0 to -1 (Fig.~\ref{fig:sweep_r}b).
With the open-end outlet, there exist a reflected $\widetilde{g}$ wave since no pressure fluctuation $\widetilde{p}$ is allowed at the outlet.
However, due to the short length of the combustion chamber compared to the acoustic wavelength, the chamber cannot sustain any cavity mode at the frequency of investigation.
Varying $\rout$ from $0$ to $-1$ is therefore used as a mean to modifiy the behaviour of the ITA loop, while keeping it isolated from cavity modes.
This shows that the acoustic waves reflected at the outlet dampen the unstable process.
The least stable eigenmode is also computed with the network model \textit{taX} \citep{EmmerJaens14}.
The eigenfrequencies quantitatively agrees and the growth rates are qualitatively close:
the damping effect of the reflected acoustic wave is enhanced in LRF (Tab.~\ref{tab:marginal_stability}).
The eigenfrequency of the ITA mode is unchanged while $\rout$ varies.
This is interpreted as the fact that changing the outlet reflection coefficient does not modify the delay time of the internal mechanism linked to the ITA mode, and therefore does not modify the structure of the ITA loop itself.
\begin{table}[ht]
    \centering
    \begin{tabular}{c|c|c}
         & LRF & \textit{taX} \\
         \hline
    $\rout$ & -0.36 & -0.44
    \end{tabular}
    \caption{Marginally stable outlet reflection coefficient $\rout$.}
    \label{tab:marginal_stability}
\end{table}

\begin{figure}[ht]
        \centering
        \includegraphics[width=0.62\linewidth]{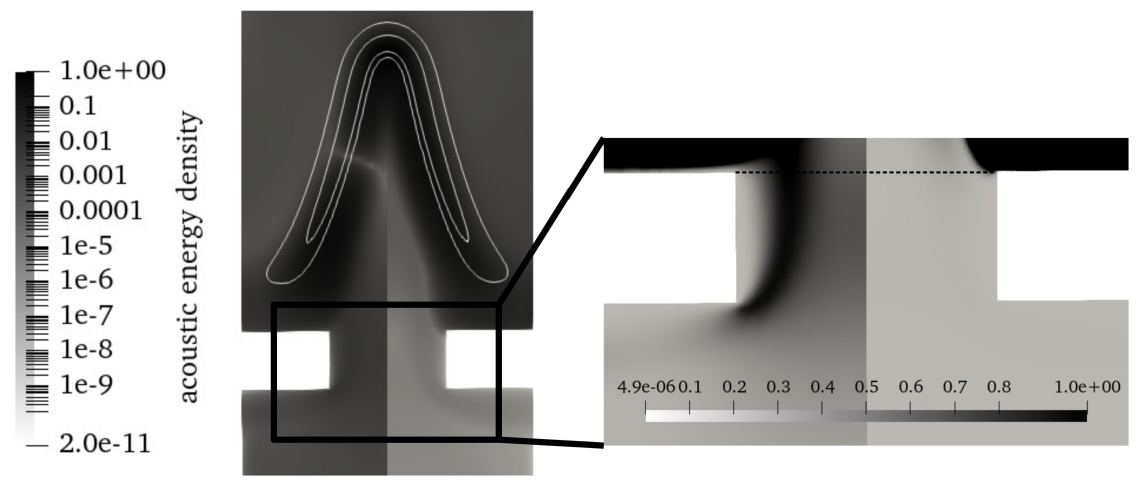}
        \caption{Acoustic energy density (Eq.~\ref{eq:acoustic_energy_density}) in logscale for (left half) $\rout=0$ and (right half) $\rout=-1$, integrated over one period.
        Zoom in natural scale normalized to the maximum along the dashed line, for both halves.}
        \label{fig:acoustic_energy}
\end{figure}

\begin{figure}[ht]
    \centering
%    \includegraphics[width=.8\linewidth]{Figs/omegaLeft.pdf}
%\caption{(left half) $\omega'^\dag$ for $\rout=0$ and (right half) $\rout=-1$ at various phase angle.}
    \sidesubfloat[]{\includegraphics[width=.2\linewidth]{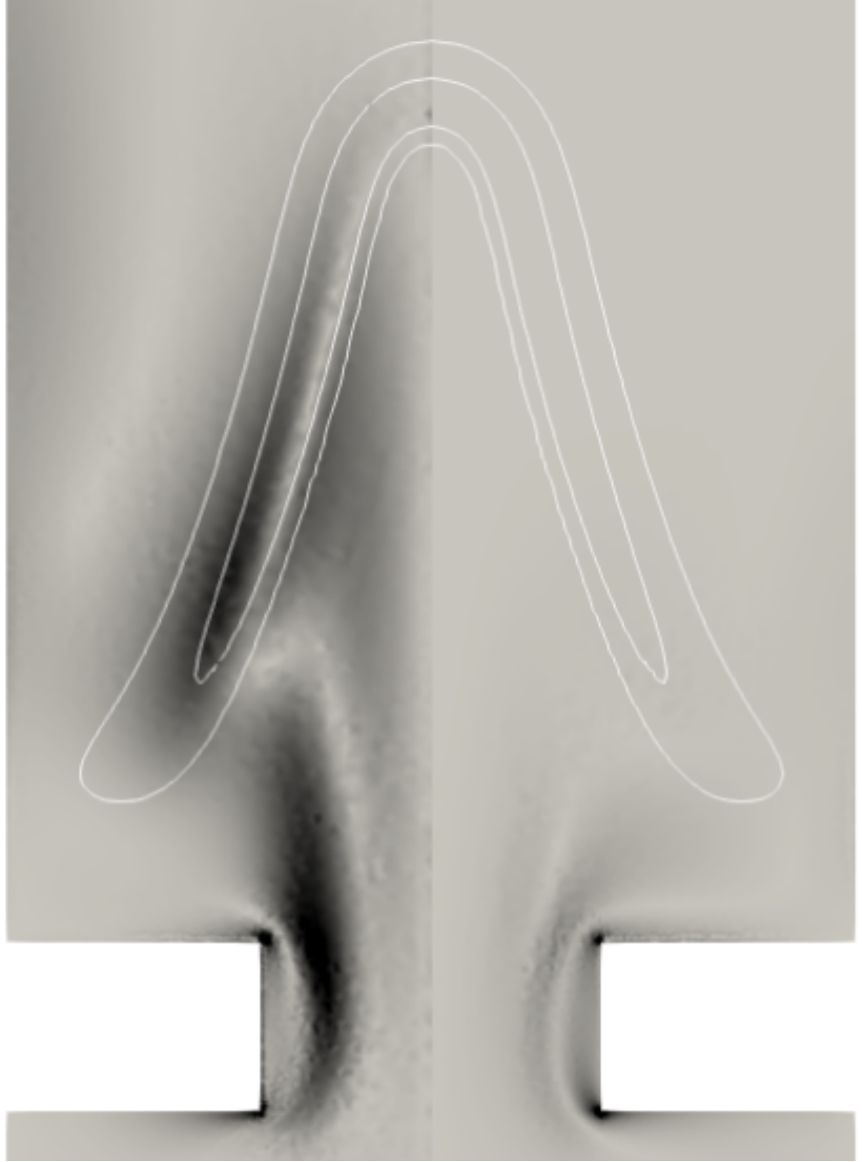}}
    \sidesubfloat[]{\includegraphics[width=.2\linewidth]{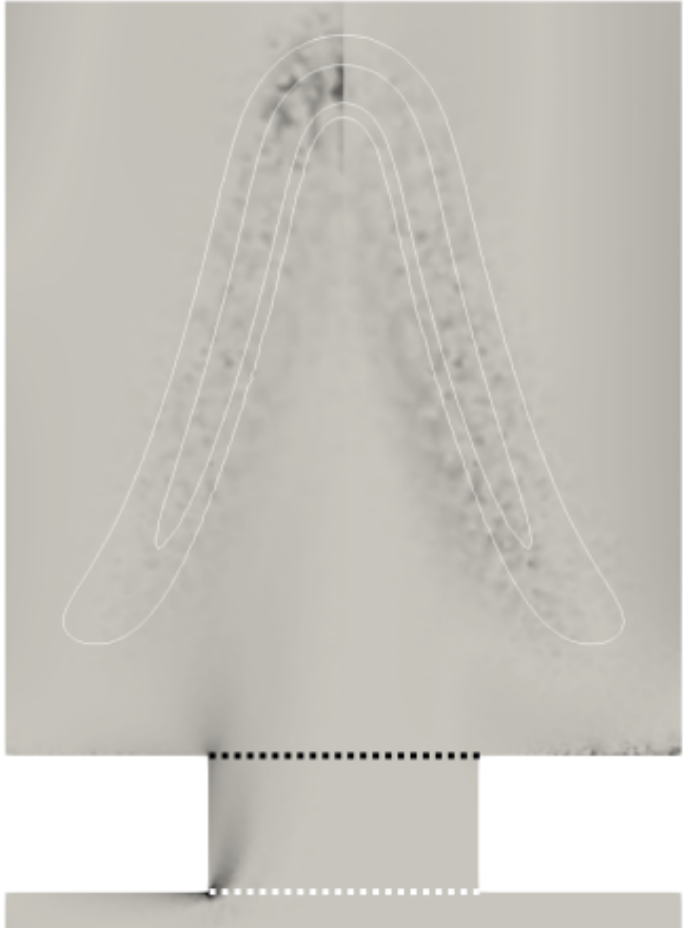}}
    \caption{(a) Absolute value of $\widetilde{\omega}^\dag$ for $\rout=0$ (left half) and $\rout=-1$ (right half). (b) Absolute value of $\widetilde{\omega}$ for (left half) $\rout=0$ and (right half) $\rout=-1$.}
\label{fig:omegaLeft}
\end{figure}

Eigenvectors corresponding to the least stable eigenmodes of Eq.~\eqref{eq:evp} are normalised by Chu's energy \citep{Chu65}, which is a relevant norm for compressible non-isentropic flows:
\begin{equation}
    E_\text{Chu}=\int_\Omega
    \dfrac{1}{2}\bar{\rho}u'^2 
    + \dfrac{1}{2} \dfrac{\bar{c}^2 p'^2}{\gamma\bar{\rho}}
    + \dfrac{1}{2} \dfrac{\Bar{\rho}\Bar{C}_vT'^2}{\Bar{T}} d\Omega.
\end{equation}
The acoustic energy density computed from \citep{Chu65}
\begin{equation}
        e_\text{ac}'=\dfrac{1}{2} \dfrac{\bar{c}^2\rho'^2}{\bar{\rho}} + \dfrac{1}{2}\bar{\rho}u'^2,
        \label{eq:acoustic_energy_density}
\end{equation}
shows that acoustics penetrates within the slit and the plenum in the non-reflective (i.e. unstable) case, but not in the reflective (i.e. stable) case (Fig.~\ref{fig:acoustic_energy}).
In details, the acoustic energy clusters in the outer region of the boundary layer within the slit in the unstable case, whereas this structure is completely absent from the stable case (zoom in Fig.~\ref{fig:acoustic_energy}).
This observation is confirmed by computing the acoustic flux
\begin{equation}
\Phi_\text{ac}=-\dfrac{1}{S}\int_{T_\text{cyc}}\int_S p' u'\cdot n \: dS \:dt, \quad\text{with}\; n \; \text{the normal vector,}
\end{equation}
entering the slit, i.e. through the black dashed line in Fig.~\ref{fig:acoustic_energy}, given in Tab.~\ref{tab:acoustic_flux}.
Since both the destabilisation of the ITA loop and the penetration of acoustic flux through the slit are obtained simultaneously for a variation on $\rout$, we conclude that penetration of the acoustic flux through the slit can be part of the ITA feedback loop.
However, that does not suffice to explain why the ITA feedback loop turns unstable.
\begin{table}[ht]
    \centering
    \begin{tabular}{c|c|c}
        $\rout$ & -1 & 0 \\
         \hline
    $\Phi_\text{ac}$ [W] & $1\: 10^{-15}$ & $7\:10^{-4}$
    \end{tabular}
    \caption{Acoustic flux entering the slit, for $\rout=0$ and $-1$.}
    \label{tab:acoustic_flux}
\end{table}
The regions where the flow is highly receptive to perturbations are identified by the adjoint of the leading eigenmode \citep{MagriJunip13b}.
In the present case, the adjoint eigenmodes $\widetilde{\mathbf{q}}^\dag$ are the left eigenvectors of Eq.~\eqref{eq:evp} and represent the optimal initial condition.
The structure of the adjoint vorticity $\widetilde{\omega}^\dag$ displays that vorticity is highly receptive in the shear layer enclosing the boundary layer of the slit and along the flame sheet (Fig.~\ref{fig:omegaLeft}a).
A potential explanation of the higher sensitivity along the flame sheet for $\rout=0$ than for $\rout=-1$ is that the eigenfrequency of the ITA eigenmode is higher for $\rout=0$ (Fig.~\ref{fig:sweep_r}b).
A higher eigenfrequency implies a shorter characteristic length of the wrinkling along the flame, therefore a more curved flame sheet that will produced more vorticity when perturbed.
The high sensitivity of vorticity in the boundary layer of the slit is consistent with general results on boundary layers, which are known to be prone to vorticity production.
Impinging perturbations, such as acoustic waves, trigger some vorticity fluctuations in this region.
The comparison between the unstable and the stable case shows that more vorticity is generated in that region in the unstable case than in the stable case (Tab.~\ref{tab:vorticity_flux}).
This vorticity production is quantified by the comparison of the perturbed vorticity between the input and the output of the slit, i.e. the white and black dotted line in Fig.~\ref{fig:omegaLeft}b: in the unstable case the vorticity fluctuation is amplified twice more throughout the slit than in the stable case.
This highly receptive region lies within the slit, and not only downstream from the edge of the slit where vorticity is shed to the combustion chamber, which corresponds to the region where acoustic energy density penetrates in the unstable case.

This analysis leads to the interpretation that the ITA loop is closed by vorticity perturbations generated \textit{within} the slit, and not only in the shear layer downstream from the slit, potentially by impinging acoustic waves penetrating the slit.
This vorticity is then convected to the base of the flame and initiates a wrinkling that produces heat release rate fluctuations.
In the present case, it is unlikely that the vorticity is generated by the baroclinic term since pressure gradients are small -- the entire system is globally acoustically compact -- and density fluctuations remain small outside the flame region.
Therefore vorticity perturbations are most likely generated by axial velocity fluctuations impinging on the non-uniform base-flow vorticity.
The unstable behaviour of the ITA loop relies on the penetration of acoustic waves within the slit and the plenum.
However, the reason why the acoustic feedback does not penetrate the slit in the stable case remains to be clearly explained.
A destructive interference can be at play in the reflective case, with $\hat{g}$ waves directly emitted from the flame and $\hat{g}$ waves reflected at the oulet cancelling one another. Note, due to the short length of the downstream chamber, both waves are almost perfectly out of phase for an open end ($\rout=-1$).
\begin{table}[ht]
    \centering
    \begin{tabular}{c|c|c}
        $\rout$ & -1 & 0 \\
         \hline
    $\int_{T_\text{cyc}}\int_{S_\text{out}} \Bar{u}_x|\omega_0'|dS/\int_{S_\text{in}} \Bar{u}_x|\omega_0'|dS\: dt$ & $7.7$ & $15.0$
    \end{tabular}
    \caption{Vorticity fluctuation $\omega_0'$ amplification for $\rout=0$ and $-1$.}
    \label{tab:vorticity_flux}
\end{table}
\subsection{From active to passive flame}
The intensity of the flame response is now varied through the prefactor $\Gamma$ in Eq.~\eqref{eq:prefactor_hrr}, such as in \citep{orchini_degenerate_2020,OrchiSilva20}.
Varying $\Gamma$ from 1 to 0 enables to investigate the behaviour of the ITA mode when the flame becomes passive, i.e. the flame does not produce unsteady heat release.
As expected, the growth rate of the ITA mode decreases with $\Gamma$ (Fig.~\ref{fig:sweep_r}c), and the heat release becomes steady.
Indeed, the flame does not produce anymore volume fluctuation, and therefore stops generating acoustic waves.
The most prominent feature of this parameter variation is the sudden change in the leading eigenvalue trajectory for $\Gamma < 0.4$.
The sensitivity of the eigenvalue is also increased, i.e. a small change $\Gamma$ induces a large change in the eigenvalue $\lambda$.
Such a change in the trend and higher sensitivity indicates a change in the physical mechanism underlying this leading eigenmode.
Due to this higher sensitivity of the leading eigenmode, it has not been possible to track the ITA eigenvalue for values $\Gamma<0.1$.
However, we hypothesize that the ITA eigenvalue tends to low frequency and negative growth rate.
This agrees with the frequency of vortex shedding behind a confined backward facing step, or a throttle in a duct.
For a laminar flow, such vortex shedding occurs at a Strouhal number, $St$ , of 0.1 to 0.01 \citep{das_unsteady_2013}.
In the present configuration,
\begin{equation}
    St=\dfrac{f(b-a)}{2\bar{u}_{x,\text{slit}}},
\end{equation}
with $\bar{u}_{x,\text{slit}}$ being the average axial velocity in the slit, and $f$ the frequency, vortex shedding frequency is expected to occur at 6Hz to 60Hz.
Such an observation advocates for a transition of the ITA loop to a purely hydrodynamic mode when the flame becomes passive, therefore becoming an open-loop since the feedback induced by the heat release rate vanishes.
The high sensitivity of vorticity in the boundary layers of the slit (Fig.~\ref{fig:omegaLeft}a) also support that hypothesis.
If confirmed, the main implication of this hypothesis is that vortex shedding from the boundary layer is recognised as part of the ITA feedback loop.
As a consequence, all parameters influencing the boundary layer of the slit would impact the ITA feedback loop: hole thickness, shape of the corners, etc.
Such a result is of importance since control strategies for ITA modes remain an open question and the response of ITA modes towards classical thermoacoustic control strategies can be rather counter-intuitive \citep{Silva23}. 

\section{Conclusion}\label{sec:conclusion}
In this study, the unsteady behaviour of the ITA loop as encountered for a laminar slit flame is found to be correlated to the penetration of acoustic flux through the slit. The loop is closed by vorticity generated in the boundary layers of the slit.
This acoustic flux stems from impinging acoustic waves generated by the unsteady heat release rate.
These results are obtained from a stability analysis of the holistic Linearized Reactive Flow model, that accounts for all peculiarities of the flow and does not need an external flame model (i.e. \textit{ad-hoc} FTF).
In the studied configuration, the combustion chamber is kept short to avoid any cavity mode developing, therefore, isolating the ITA loop.
The behaviour of the ITA eigenmode is studied for varying outlet reflection coefficient and heat release rate prefactor, rendering the flame passive.
The adjoint eigenmode shows a high sensitivity of vorticity in the boundary layer of the slit, meaning that any perturbation of the flow here would produce vorticity.
In particular, acoustic perturbations travelling upstream from the flame and penetrating the slit will produce vorticity in the slit.
This vorticity will then be advected by the bulk flow to the base of the flame where it initiates a wrinkling of the flame sheet that will propagate up to the flame tip and induce a heat release rate fluctuation.
Variation of the outlet reflection coefficient shows that the ITA loop turns unstable when a significant part of the acoustic flux penetrates the slit.
However, the reason why the acoustic flux does not penetrate the slit when zero pressure fluctuation is enforced at the outlet ($\rout=-1$) remains unexplained.
Based on the heat release rate prefactor variation, we hypothesize that the ITA loop transitions to a purely hydrodynamic open-loop mode for a passive flame.
This is backed by the high sensitivity of vorticity in the boundary layer of the slit.
Such an outcome would be of importance in the perspective of controlling thermoacoustic instabilities \citep{Silva23}.
Indeed, if this hypothesis is confirmed, it would imply that the characteristics of the slit -- aspect ratio, shape of corners, roughness of walls, etc. -- affect the ITA loop.
The linear analysis framework could here provide tools to optimally choose these parameters, so as to dampen the unstable ITA mode when it arises.
However, it was not possible to track the ITA eigenmode down to a complete passive flame because of the too large sensitivity of the spectrum.
Such an issue can be circumvented \textit{via} an efficient eigenvalue tracking algorithm.

Although the purpose of this study is to isolate the ITA loop, realistic combustion chambers also support cavity modes, and the resulting thermoacoustic oscillations are the interplay of both, and not simply a superposition of them.
The present study should therefore be extended to include these cavity modes.

%\begin{thebibliography}{9}
%\bibliographystyle{alpha}
\bibliographystyle{apalike}
\bibliography{ITA} 
%\printbibliography
%\bibliography{ITA}
%\end{thebibliography}
%\bibliography{ITA}
\end{document}